\documentclass[pre,twocolumn,showpacs,aps,floats,floatfix,superscriptaddress]{revtex4}

\usepackage{graphicx}
\begin{document}

\newcommand{\avk}{\langle k \rangle}
\newcommand{\fluck}{\langle k^2 \rangle}

\title{Arrival Time Statistics in Global Disease Spread}

\author{Aur\'elien Gautreau}
\affiliation{LPT, CNRS, UMR 8627, and 
Univ Paris-Sud, Orsay, F-91405  (France)}
\author{Alain Barrat} 
\affiliation{LPT, CNRS, UMR 8627, and 
Univ Paris-Sud, Orsay, F-91405  (France)}
\affiliation{Complex Networks Lagrange Laboratory, ISI Foundation, 
Turin, Italy}\author{Marc Barth\'elemy}
\affiliation{CEA-Centre d'Etudes de
Bruy{\`e}res-le-Ch{\^a}tel, D\'epartement de Physique Th\'eorique et
Appliqu\'ee BP12, 91680 Bruy\`eres-Le-Ch\^atel, France}
\date{\today} \widetext
\begin{abstract}
  
  Metapopulation models describing cities with different populations
  coupled by the travel of individuals are of great importance in the
  understanding of disease spread on a large scale. An important
  example is the Rvachev-Longini model [{\it Math. Biosci.} {\bf 75},
  3-22 (1985)] which is widely used in computational epidemiology. Few
  analytical results are however available and in particular little is
  known about paths followed by epidemics and
  disease arrival times. We study 
  the arrival time of a disease in a city as a function of the
  starting seed of the epidemics. We propose an analytical
  Ansatz, test it in the case of a spreading on the world wide air
  transportation network, and show that it predicts accurately the
  arrival order of a disease in world-wide cities.

\end{abstract}



\maketitle 


In modern societies, individuals can easily travel over a wide range
of spatial and temporal scales. The interconnections of areas and
populations through various means of transport have important effects
on the geographical spread of epidemics. In particular, the structure
and the different complexity levels of the air-transportation network
are responsible for the heterogeneous and seemingly erratic outbreak
patterns observed in the worldwide propagation of
diseases~\cite{Colizza:2005} as recently documented for
SARS~\cite{SARS,Hufnagel:2004}.  In order to describe such a complex
phenomenon and to obtain powerful numerical forecasting tools,
different levels of description are possible, ranging from a simple
global mean-field to detailed agent-based simulations
\cite{May:1992,Hethcote:1984,Keeling:1999,
Pastor:2001,May:2001,Ferguson:2003,Eubank:2004} that recreate entire
populations and their dynamics at the scale of the single
individual~\cite{Eubank:2004}.

At large scale, such as the world-wide level, a very important class
of models in modern epidemiology are the so-called metapopulation
models which use a description at two levels by dividing the global
population into interconnected subpopulations.  Within each
subpopulation, a mean-field like model of epidemic spreading is used,
while the spread from one subpopulation to another is due to the
travel of individuals. Agents of each subpopulation can be in various
states (healthy, infectious, recovered...), change state by
contact with other agents, and diffuse on the
transportation network between subpopulations. Metapopulation models
can thus be considered as reaction-diffusion processes, which opens
very interesting perspectives and issues~\cite{Colizza:2007} within
the global framework of dynamical phenomena occurring on complex
networks
\cite{Barabasi:2000,Doro:2003,Pastorbook:2003,Boccaletti:2006}.  For
the description of worldwide epidemic spreading, the subpopulations
are cities connected by a transportation network in which links
correspond to the existence of passenger flows described by the
worldwide air-transportation network (WAN). The WAN represents a major
channel for the worldwide spread of infectious diseases
\cite{Hufnagel:2004,Colizza:2005} and its complex, heterogeneous
features at various levels (degree distribution, traffic, populations)
have recently been characterized~\cite{Amaral:2004,Barrat:2004}.

In this Letter, we focus, in the framework of such metapopulation models, on
the issue of the arrival time in a city of the first infectious individual.
In particular, we study how this time depends on the origin of the
disease and on the network characteristics.
This problem is more complex than the one of random walks on complex
networks \cite{Rieger:2003}, since the number of infectious individuals
diffusing on the network is constantly evolving due to the inner-city epidemic
dynamics. We also note that references \cite{bart:2005,bart:2006} were also
concerned with the arrival time problem for an epidemic spreading on a complex
network, but in a different framework: each network node was an individual
(susceptible or infectious), while in our case each node
represents a whole subpopulation.  After the precise definition of the model,
we will first consider the simple case of a one-dimensional topology for the
transportation network in order to gain analytical insights into this problem.
This will allow us to propose an analytical form for the arrival time in
arbitrary networks. We then test this form in the case of the WAN, by
simulating numerically a stochastic spreading phenomenon on the network, and
show that we can indeed predict with a good accuracy the spreading phenomenon
and the arrival order of a disease in various cities at a world-wide level.


While the precise
model describing the epidemic spreading at the subpopulation level
could be refined at will in order to describe a particular disease, we
are here interested in generic and fundamental aspects of the
metapopulation modeling approach. We therefore restrict our
study to a simple SI disease model in which individuals are either
healthy (susceptible, S) or can become infectious (I)
if in contact with an infectious individual. 
The Rvachev-Longini SI model \cite{Longini:1985} describes the evolution of
the number of infectious $I_i(t)$ individuals (and also of $S_i(t)$) 
 in each city $i$
through
\begin{equation}
\partial_tI_i=K(\{X_i\})+\Omega(\{I_j\}) \ ,
\end{equation}
where the first term $K$ of the right hand side describes the (epidemic)
reaction process inside each subpopulation (city), due to the interaction of
individuals in the various possible states. In our case $X \in \{S,I\}$ (we
have checked that more involved models, such as SIS or SIR, give consistent
results \cite{inprep}) and the standard homogeneous mixing assumption in each
city gives~\cite{May:1992}: $K(\{X_i\})=\lambda I_i(N_i-I_i)/N_i$, where $N_i$
is the population of city $i$ and $\lambda$ the spreading rate.  The second
term $\Omega$ represents the evolution due to the arrival or departure of
infectious individuals from or to other cities and is determined by passenger
flows on the transportation network. This model therefore considers a
simplified mechanistic approach with a widely used markovian assumption in
which individuals are not labeled according to their original subpopulation,
and at each time step the same traveling probability applies to all
individuals in the subpopulation, without memory of their origin
\cite{Longini:1985,Hufnagel:2004,Colizza:2005}. Denoting by $w_{ij}$ the
average number of passengers traveling from $i$ to $j$ per unit of time
($w_{ij}=0$ if there is no direct connection), the probability per unit time
that an individual travels from city $i$ to city $j$ is then given by
$w_{ij}/N_i$. The full metapopulation model is therefore described by
\begin{equation}
\partial_t I_i=\lambda I_i(t)\frac{N_i-I_i(t)}{N_i}+
\sum_j\frac{w_{ji}}{N_j}I_j-\sum_j\frac{w_{ij}}{N_i}I_i \ .
\label{RLcont}
\end{equation}
This original formulation considers only expectation values, which can
take continuous values, so that ``fractions'' of infectious
individuals can travel and infect neighboring cities arbitrarily fast
\footnote{A numerical integration of (\ref{RLcont}) leads to an infection of
  all cities at time $0^+$.}. To investigate arrival times, one therefore
needs to take into account the inherent stochasticity of the
spreading. We thus consider in all our numerical
simulations the stochastic generalization described in
\cite{Colizza:2005,Hufnagel:2004} where the number of
individuals traveling on each connection is an integer variable randomly
extracted at each time step of length $\Delta t$, with average $\Delta t
w_{ij} I_i/N_i$ (in the numerical simulations we will use $\Delta t=1\; {\rm
  day}$); for simplicity we keep the endogenous growth deterministic since we
are mainly concerned with the effect of travel, but we have checked that
inclusion of stochastic effects as in \cite{Colizza:2005} do not change our
results \cite{inprep}. Note that in real cases such as the WAN, most weights
are symmetric ($w_{ij}= w_{ji}$)~\cite{Barrat:2004} but the probabilities of
travel from one city to another are not since they depend on the
populations of the various cities: the travel effectively occurs as a random
diffusion with non-symmetric rates on the transportation network.
The topological distance thus does not contain all the information needed
to characterize such a process. Moreover, since most transportation networks
are small-world networks, many cities lie at the same topological distance
from a given seed, but will potentially be reached at very different times.

\begin{figure}[htb]
  \vspace*{0.5cm}
   \centering
   \includegraphics[angle=0,scale=.25]{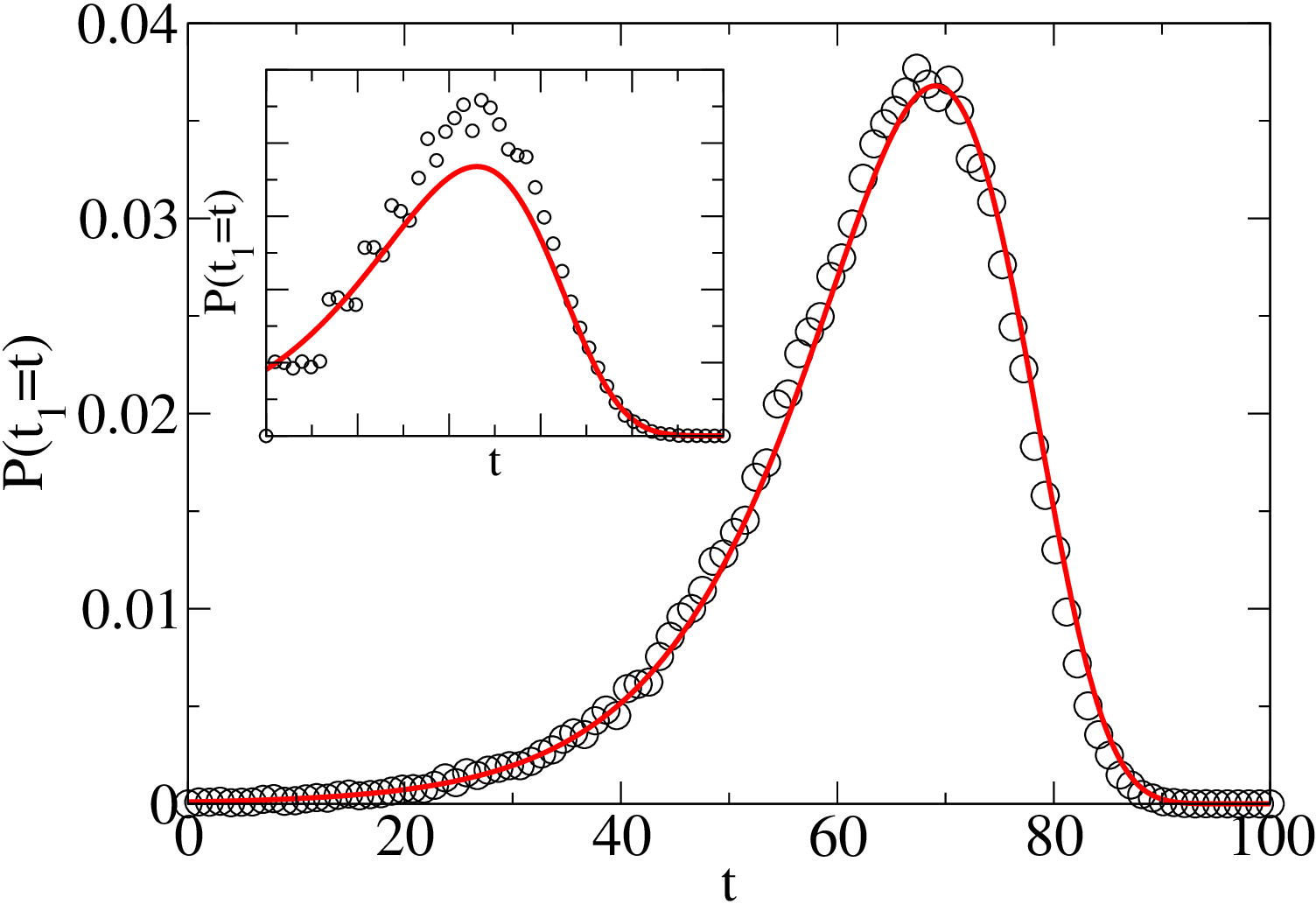}
   \caption{Two cities model. Arrival time $t_1$ distribution computed
     by numerical simulation and compared with the result of
     Eq.~(\ref{eq:distrfin}) for
     $\frac{w}{N_0\lambda}=10^{-2}$ (line). \textit{Inset}: 
     Same for $\frac{w}{N_0\lambda}=10^{-1}$.}
   \label{fig:gumbel}
\end{figure}

Before turning to numerical simulations of the described model, we present an
analytical approach to the determination of arrival times.  Let us first
consider the simple case of two cities ($0$ and $1$), with populations $N_0$,
$N_1$ which are connected by a passenger flux $w_{01}=w$.  We assume that at
$t=0$, there are $I^0=1$ infectious people in the city $0$.  Let us first
consider that the travel events occur as instantaneous jumps of probability
$p=\frac{w}{N_0}\Delta t$, at discretized times, in units of $\Delta t$. The
probability that the time of arrival $t_1$ of the epidemic in the city $1$ is
equal to $t=n\Delta t$ is then
\begin{equation}
P_{d}(t_1=n\Delta t) =
\left[1-(1-p)^{I^0(n \Delta t)}\right]
\prod_{i=1}^{n-1} (1-p)^{I^0(i\Delta t)} \ .
\label{eq:probaI}
\end{equation}
In order to obtain the density probability $P(t)$ of the arrival time
in the city $1$ we consider the limit $\Delta t \to 0$, using the
following assumptions: (i) $I^0(t)\ll N$, which is realistic for usual
diseases, in which only small fractions of the population are
infectious; (ii) the continuous limit for $I^0(t)$ can be used which
reads $\frac{1}{\lambda}\ll \langle t_1\rangle$. Within these
assumptions, we obtain
\begin{equation}
P(t)dt=\frac{w}{N_0}e^{\lambda t-\frac{w}{N_0\lambda}e^{\lambda t}}dt
\label{eq:distrfin}
\end{equation}
(the last assumption then reads $1\ll \ln(\frac{N_0\lambda}{w})$). We
recognize in (\ref{eq:distrfin}) a Gumbel distribution with average
$\langle t_1\rangle
=\frac{1}{\lambda}[\ln(\frac{N_0\lambda}{w})-\gamma]$, where $\gamma$
is the Euler constant. The variance is
$Var(t_1)=\frac{\pi}{\sqrt{6}\lambda}$ and does not depend of
$\frac{w}{N_0}$ (the non physical contribution of the negative values
of $t$ in the distribution has to be negligible which is satisfied if
$\int_{-\infty}^0P(t)dt=\frac{w}{N_0\lambda}\ll 1$). Within these
assumptions, we obtain a good agreement between results of
numerical simulations using discretized travel events
\cite{Colizza:2005} and the theory which uses continuous 
approximations (see Fig.~\ref{fig:gumbel}).

\begin{figure}[htb]
  \vspace*{0.1cm}
   \centering
   \includegraphics[angle=0,scale=.35]{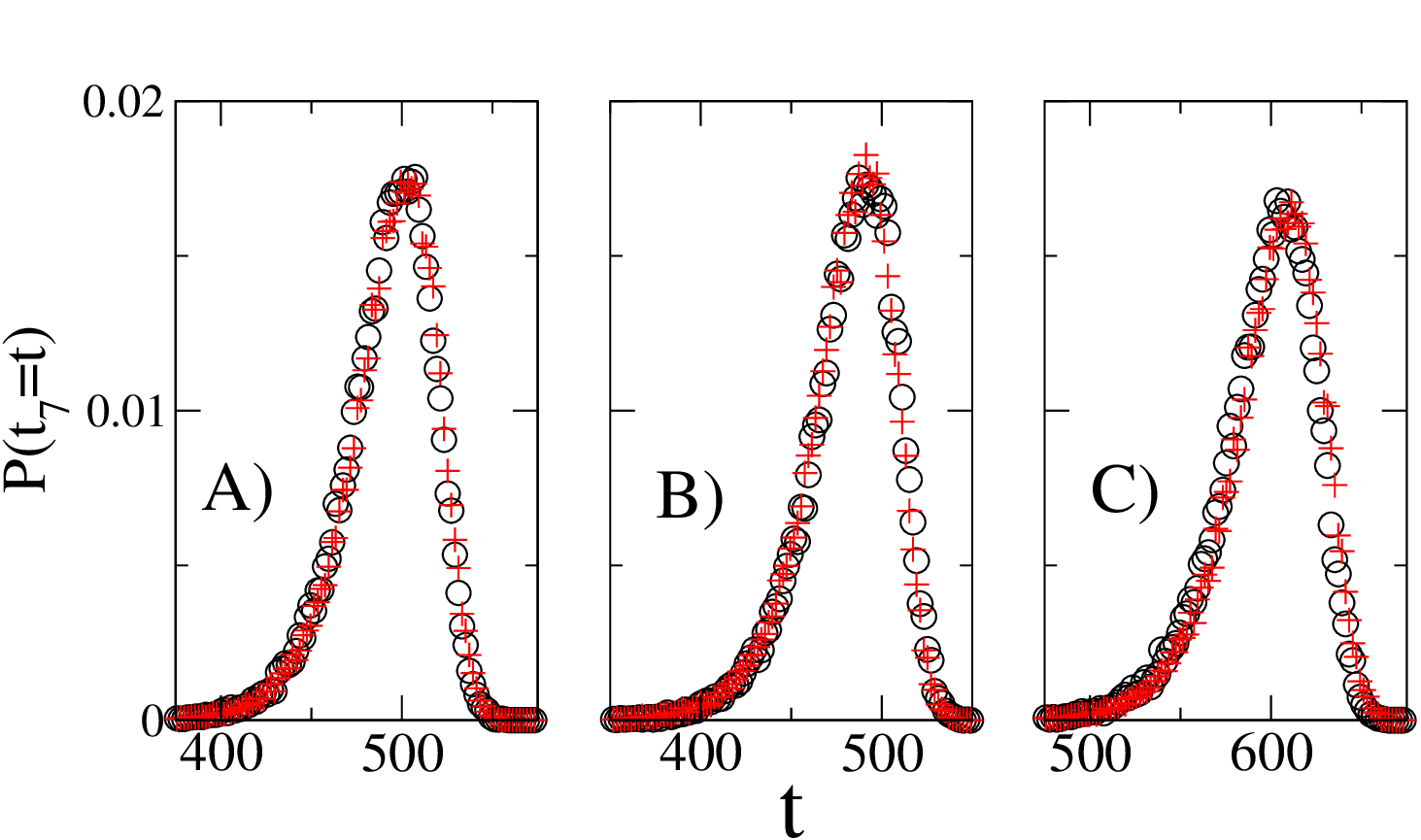}
   \caption{ (A-C) Arrival time
     distribution on a line at city $\#$7, from numerical
     simulations for a fixed random set of populations $\{N_i\}$ and
     weights $\{w_i\}$ (Black circles). Red crosses:  
     distributions for (A) uniform travel $w_i=\bar{w}$,
     and populations $\{N_i\}$; (B) uniform populations $N_i=\bar{N}$,
     and weights $\{w_i\}$; (C) uniform populations $N_i=\bar{N}$ and
     weights $w_i=\bar{w}$. We use a small value of $n$ since most real
complex networks are small-world: any node lies at a small distance from
     the seed.}
   \label{fig:sym}
 \end{figure}

We now consider the case of a one-dimensional line of cities connected by
passenger fluxes of random intensity.  We assume that the spreading process
starts at city $0$ and we denote by $t_n$ the arrival time in the city $n$.
The quantities having the same unit as $t_n$ are $1/\lambda$ and $N_i/w_i$,
where $w_i$ is the number of passengers traveling from $i$ to $i+1$ per unit
time. Dimensional analysis then implies that the probability distribution of
the adimensional quantity $\lambda t_n$ must be a function of the other
adimensional quantities which are the $w_i/(lambda N_i)$: $P(\lambda
t_n)=G_n(\lambda t_n,\{\frac{w_i}{N_i\lambda}\})$, where $G_n$ is an unknown
function. One can write $t_n$ as a sum of random variables,
$\Delta_i=t_{i}-t_{i-1}$ which are however correlated since each local
infection process depends on the history of the epidemics in all previously
infected cities. While a complete study of $P(\lambda t_n)$ is left for future
work \cite{inprep}, numerical simulations of the spreading show (Fig.
\ref{fig:sym}) that it obeys important invariance properties. For
heterogeneous populations and travels ($w_i$ and $N_i$ are distributed
uniformly in $[10,2000]$ and $[10^5,2.10^7]$, respectively), the whole
distribution is invariant when one replaces (i) all the random weights by
their geometrical mean $\overline{w}=(\prod_{i=0}^{n-1}w_i)^{1/n}$; (ii) all
the random populations by their geometrical mean
$\overline{N}=(\prod_{i=0}^{n-1}N_i)^{1/n}$; (iii) all weights by
$\overline{w}$ and all populations by $\overline{N}$.  The ratios of the
average times for these different sets stay very close
to $1$, with deviations at most of the order of $5\%$.

The average arrival time can thus be written as $\lambda \langle
t_n\rangle =F(\{\frac{w_i}{N_i\lambda}\})$ where $F(x_1,\dots,x_n)$ is
a symmetric function of its variables which depends only on the
product $\prod x_i$, and such that $\langle t_1\rangle$ is the average
of the Gumbel distribution (\ref{eq:distrfin}). This leads to
the following Ansatz
\begin{equation}
\lambda \langle t_n\rangle \approx 
\chi(n) \equiv \ln\left[\prod_{i=0}^{n-1}\frac{N_i\lambda e^{-\gamma}}{w_i}\right] \ .
\label{eq:chi}
\end{equation}

\begin{figure}[htb]
  \vspace*{0.5cm}
   \centering
   \includegraphics[angle=0,scale=.25]{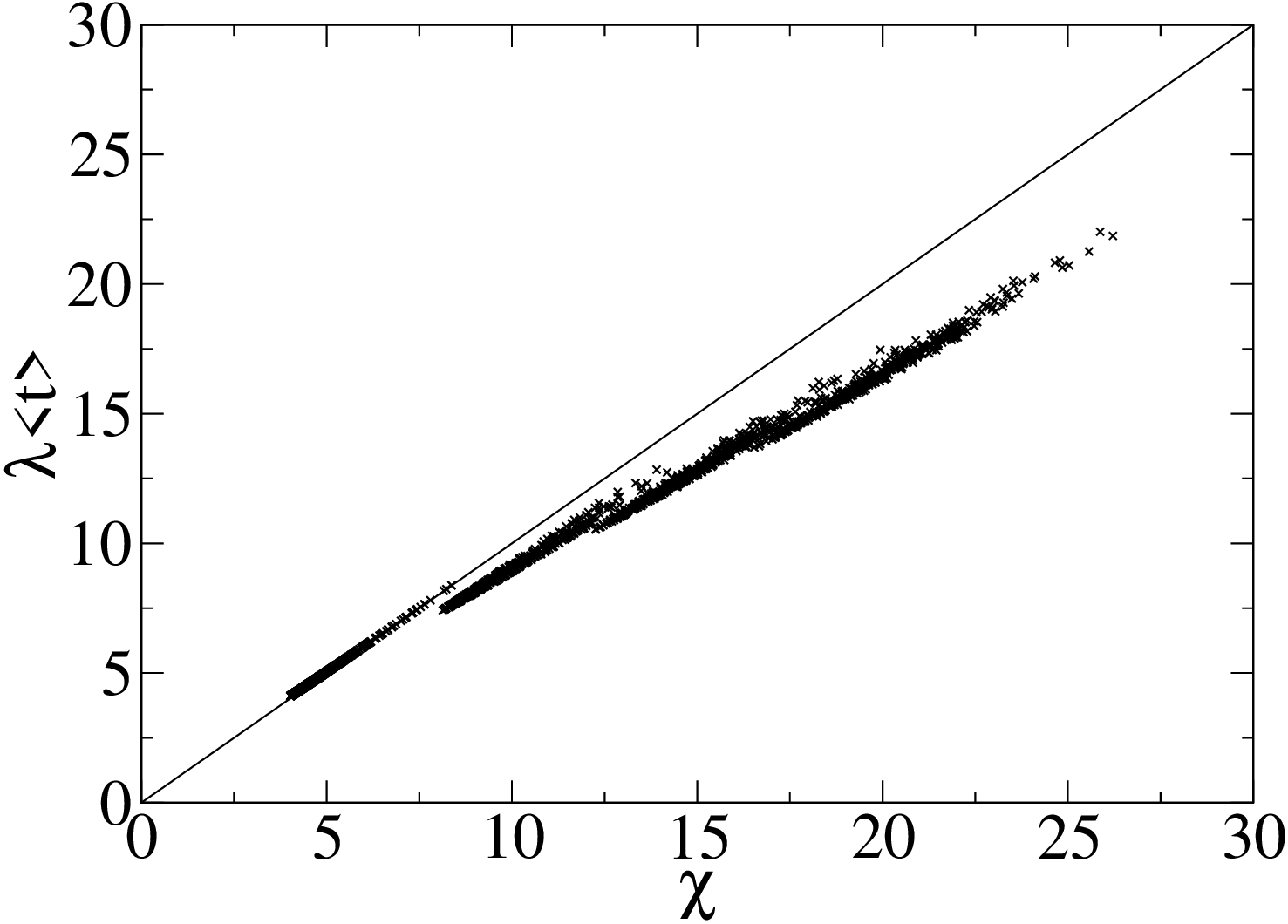}
   \caption{$\lambda\langle t_n\rangle$ vs. $\chi(n)$ for $5$ 
     cities connected on a line, with $100$ different random sets
     $\{w_i,N_i\}$. Each point is an average over $1,000$ epidemics
     for each realization of the $w$'s. }
   \label{fig:chi}
\end{figure} 
Figure \ref{fig:chi} shows that the average arrival time in a city is
indeed determined by $\chi$ to a very good extent (while the arrival
time at a given topological distance from the seed can vary a lot).
More quantitatively, $\chi$ is approximately proportional to
$\lambda\langle t\rangle$, which it slightly overestimates since we
neglect the flow of infectious individuals 
from $n-2$ to $n-1$ with respect to the
endogenous increase of $I_{n-1}$ during $[t_{n-1};t_n]$
\cite{inprep}.

We now consider a generic transportation network between the cities.  The
quantity (\ref{eq:chi}) can easily be computed on any path of length $n$ on
the network.  While the spread can a priori follow multiple paths from one
city to another, we can reasonably assume that the most probable path is the
one which minimizes the value of $\chi$ computed on it, leading to the
smallest arrival time possible (a more refined Ansatz
taking into account multiple paths does not lead to strong
differences in the final results~\cite{inprep}). We thus obtain the following
Ansatz for the arrival time at a city $t$ of a disease
starting at node $s$
\begin{equation}
\chi(s,t)=\min_{\{P_{st}\}}\sum_{(k,l) \in P_{st}}
\left[\ln\left(\frac{N_k\lambda}{w_{kl}}\right)-\gamma\right]
\label{eq:ansatz}
\end{equation}
where $\{P_{sj}\}$ is the set of all possible paths connecting $s$ to
$t$, and the sum is over the links $(k,l)$ on the paths. In other
terms, we have introduced a new (non symmetric)
weight $\ln (N_i\lambda/w_{ij})-\gamma$
 on each oriented link $(i,j)$ of the network.


\begin{figure}[!t]
  \vspace*{0.5cm}
   \centering
   \includegraphics[angle=0,scale=.25]{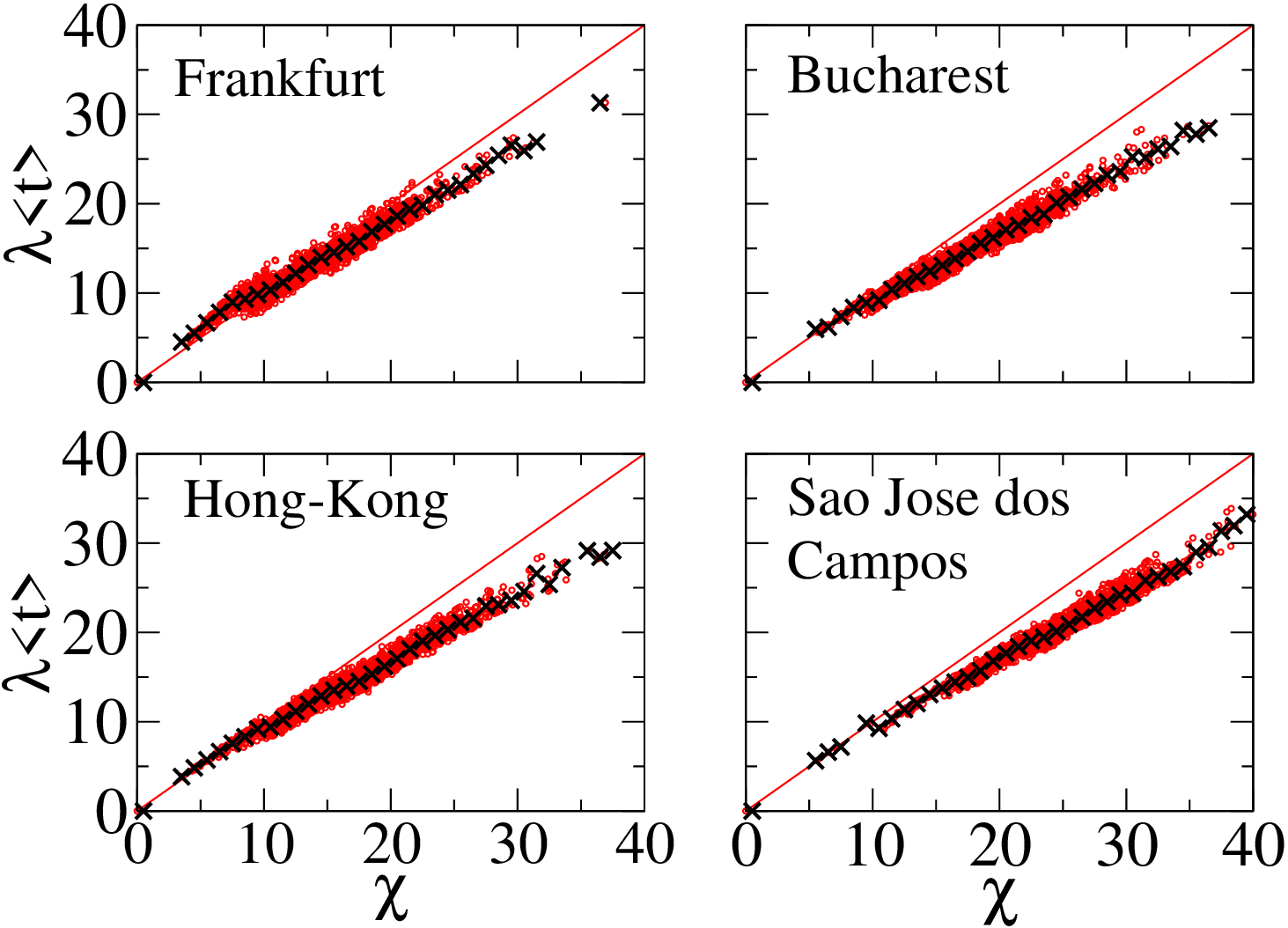}
   \caption{(Color online) $\lambda \langle t\rangle$ versus $\chi$ on
     the WAN for diseases starting in different cities (whose name
is specified in each graph).
 Each red circle corresponds to
     a city and averages are done over $1,000$ realizations of the
     spreading. Crosses are an average over cities with the same
     $\chi$. When the initial seed is a hub, the average arrival time
     is larger than $\chi$  in the first reached cities,
     due to the multiplicity of possible paths \cite{inprep}.}
   \label{fig:WAN}
 \end{figure} 
 
\begin{figure}[!t]
 \vspace*{0.5cm}
   \centering
   \includegraphics[angle=0,scale=.25]{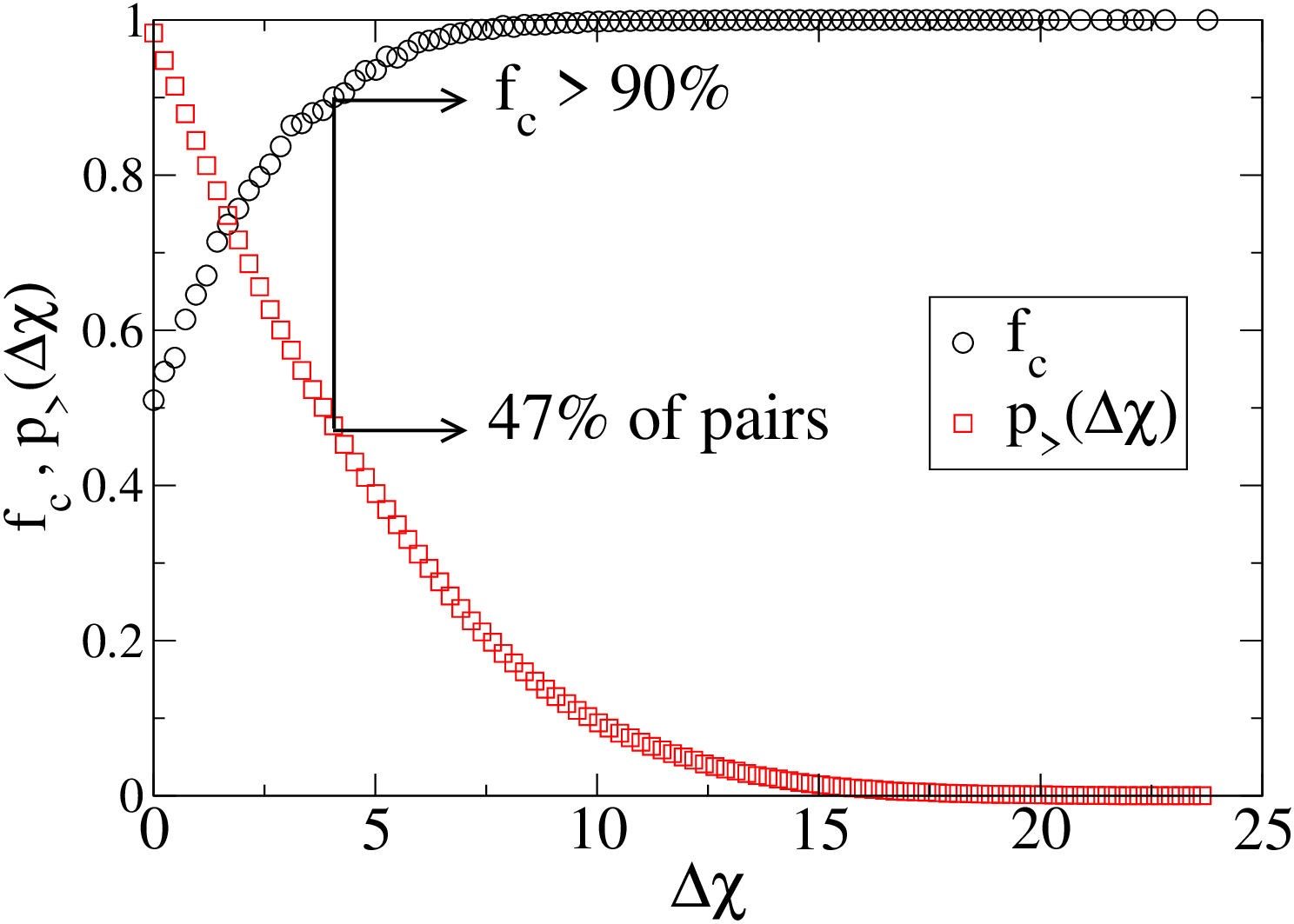}
   \caption{Fraction of couples of nodes correctly ranked as a function of
     their $\Delta\chi$ (circles), in each realization of the spread,
and cumulative distribution (squares) of
     the values of $\Delta\chi$ (i.e., fraction of couples of cities $(i,j)$
     with $\Delta\chi(i,j) > \Delta\chi$).  }
   \label{fig:correct}
\end{figure}

 We have simulated, using the model developed in \cite{Colizza:2005}, and
 summarized above, a spreading phenomenon on a subnetwork of the WAN, composed
 of the $2,400$ nodes for which the populations are larger than $10,000$
 inhabitants and which corresponds to $98\%$ of the total traffic
 \footnote{Similar results are obtained for simulations on various network
   models, but we present results for the WAN which contains additional
   correlations which may not be present in many models \cite{Barrat:2004}.}.
 The arrival times are computed by solving numerically the equations of the
 Rvachev-Longini model with discretized random travel events, and averaging
 over $1,000$ realizations of the spreading with the same seed (one infectious
 individual in a given city). Figure~\ref{fig:WAN} shows the obtained values
 of $\lambda \langle t\rangle$ versus $\chi$ for various initial seeds. We
 observe that the average arrival time is indeed determined by the value of
 $\chi$ in a given city: various cities with the same $\chi$ are reached at
 the same time by the disease propagation. While $\chi$ quantitatively
 overestimates the arrival time, the two quantities are correlated strongly
 enough, in order to obtain with a good confidence the order of arrival of the
 disease in different cities.  More precisely, if we denote
 $\Delta\chi(i,j)=\mid\chi(j)-\chi(i)\mid$, we show in Fig. \ref{fig:correct}
 the probability $f_{c}(\Delta\chi)$ that the arrival times {\em in one
   realization of the spread} $t^{(i)}$ and $t^{(j)}$ follow the same order as
 given by $\chi(i)$ and $\chi(j)$ [ie. $(t^{(i)}-t^{(j)})(\chi(i)-\chi(j)) >
 0$]. In other words, $f_{c}$ is the probability that the disease arrival rank
 for the two cities $i$ and $j$ is correctly predicted by $\chi$. If
 $\Delta\chi(i,j)$ is equal to $0$, no prediction is possible and we indeed
 obtain $f_{c}(0)=0.5$. For $\Delta\chi>10$, almost all node couples are
 correctly ranked. This result has however to be weighted by the number of
 couples with such a large $\Delta\chi$. We thus plot on the same figure the
 cumulative distribution $p_>(\Delta\chi)$ of the number of couples of nodes
 with a given value of $\Delta\chi$. We see for example that approximately
 $80\%$ of the couples of cities have a $\Delta\chi>2$ and more than $70\%$ of
 these couples are correctly sorted (instead of just $50\%$ on average if no
 information is available).

From a theoretical point of view, metapopulation models go far beyond
classical random walks and deserve many further theoretical investigations.
In this Letter, we have proposed an Ansatz for the arrival time of a disease
in a city, knowing the starting point of the spread. This Ansatz is a good
approximation and predicts with accuracy the arrival order of the disease in
the different cities, even if they are at the same topological distance from
the seed \cite{inprep}. Containment strategies could use such information to
target the cities most at risk of rapid infection, and therefore
deploy limited supplies of vaccine or antivirals in an efficient way.
Further developments could include more
sophisticated compartmental or metapopulation models, and the
systematic investigation of various structures of complex networks
\cite{inprep}. Finally, it would be interesting to extend this study
to other scales, like the urban scale, where nodes are locations such as 
homes, offices or malls~\cite{Eubank:2004}.

Acknowledgments. We thank V. Colizza and A. Vespignani for discussions
and comments. A.G. and M.B. also thank the School of Informatics,
Indiana University where part of this work was performed. We are very
grateful to V. Colizza for providing the data on city populations and
we thank IATA (http://www.iata.org) for making their database
available to us.  A.G. and A.B. are partially supported by EU
contract 001907 (DELIS).




\begin{thebibliography}{99}



\bibitem{Colizza:2005} V. Colizza, A. Barrat, M. Barth\'elemy,
  A. Vespignani, Proc. Natl. Acad. Sci. (USA) {\bf 103},
  2015 (2006); Bull. Math. Biol. {\bf 68}, 1893 (2006).

\bibitem{SARS} \url{http://www.who.int/csr/sars/en} 

\bibitem{Hufnagel:2004}
L. Hufnagel, D. Brockmann and T. Geisel,
Proc. Natl. Acad. Sci. (USA) {\bf 101}, 15124 (2004).

\bibitem{May:1992} R. M. Anderson and R. M. May , {\it Infectious 
diseases in humans} (Oxford University Press, Oxford, 1992). 
 
\bibitem{Hethcote:1984} H. W. Hethcote and J. A. Yorke, 
Lect. Notes Biomath. {\bf 56} (Berlin, Springer-Verlag, 1984). 
 
\bibitem{Keeling:1999} M. J. Keeling, 
Proc. R. Soc. Lond. B {\bf 266}, 859 (1999). 
 
\bibitem{Pastor:2001} R. Pastor-Satorras and A. Vespignani,
Phys. Rev. Lett. {\bf 86}, 3200 (2001). 
 
\bibitem{May:2001} A. L. Lloyd and R. M. May,
Science  {\bf 292}, 1316 (2001). 
 
\bibitem{Ferguson:2003} N. M. Ferguson, M. J. Keeling, W. J. Edmunds, 
R. Gani, B. T. Grenfell, R. M. Anderson and S. Leach,
Nature {\bf 425}, 681 (2003).
 
\bibitem{Eubank:2004} S. Eubank, H. Guclu, V. S. Anil Kumar, 
M. V. Marathe, A. Srinivasan, Z. Toroczkai and N. Wang, 
Nature {\bf 429}, 180 (2004).

\bibitem{Colizza:2007} V. Colizza, R. Pastor-Satorras and A. Vespignani,
  Nature Physics {\bf 3}, 276 (2007).
 
\bibitem{Barabasi:2000} R. Albert and A.-L. Barab{\'a}si,
Rev. Mod. Phys. {\bf 74}, 47 (2000). 
 
\bibitem{Doro:2003} S. N. Dorogovtsev and J. F. F. Mendes, {\em Evolution of
    networks: From biological nets to the {I}nternet and {WWW}} (Oxford
  University Press, Oxford, 2003).
 
\bibitem{Pastorbook:2003} R. Pastor-Satorras and A. Vespignani, 
{\em Evolution and structure of the Internet: A statistical physics 
approach} (Cambridge University Press, Cambridge, 2003). 

\bibitem{Boccaletti:2006}
S. Boccaletti, V. Latora, Y. Moreno, M. Chavez and D.-U. Hwang,
Phys. Rep. {\bf 424}, 175 (2006).

\bibitem{Barrat:2004} A. Barrat, M. Barth{\'e}lemy,
  R. Pastor-Satorras and A. Vespignani, 
    Proc. Natl. Acad. Sci. (USA) {\bf 101}, 3747 (2004).

\bibitem{Amaral:2004} R. Guimer\`a and L. A. N. Amaral, 
Eur. Phys. J. B {\bf 38}, 381 (2004).

\bibitem{Rieger:2003} J. D. Noh and H. Rieger, Phys. Rev. Lett. {\bf
    92}, 118701 (2004).


\bibitem{bart:2005} M. Barth\'elemy, A. Barrat, R. Pastor-Satorras and 
A. Vespignani, Phys. Rev. Lett. {\bf 92}, 178701 (2004).

\bibitem{bart:2006} P. Cr\'epey, F. P. Alvarez and M. Barth\'elemy, 
Phys Rev E {\bf 73}, 046131 (2006).

\bibitem{Longini:1985} L.A. Rvachev and I.M. Longini, Mathematical 
Biosciences {\bf 75}, 3 (1985).
 

\bibitem{inprep}
A. Gautreau, A. Barrat and M. Barth\'elemy, in preparation.

\end{thebibliography}
\end{document}